# Photonic time-delayed reservoir computing based on series coupled microring resonators with high memory capacity

Yijia Li[1], Ming Li[2], Mingyi Gao[3], Chang-Ling Zou[2], Chun-Hua Dong[2], Jin Lu[5], Yali Qin[1], XiaoNiu Yang[1,4], Qi Xuan[1,4], and Hongliang Ren[1,*]

[1] *College of Information and Engineering, Zhejiang University of Technology, Hangzhou 310023, China;*
[2] *Key Laboratory of Quantum Information, CAS, University of Science and Technology of China, Hefei, Anhui 230026, China;*
[3] *School of Electronic and Information Engineering School, Soochow University, Suzhou 215006, China*
[4] *Institute of Cyberspace Security, Zhejiang University of Technology, Hangzhou 310023, China*
[5] *College of Computer Science and Technology, Zhejiang University of Technology, Hangzhou 310023, China;*

[*] *hlren@zjut.edu.cn*

**Abstract:** On-chip microring resonators (MRRs) have been proposed to construct the time-delayed reservoir computing (RC), which offers promising configurations available for computation with high scalability, high-density computing, and easy fabrication. A single MRR, however, is inadequate to supply enough memory for the computational task with diverse memory requirements. Large memory needs are met by the MRR with optical feedback waveguide, but at the expense of its large footprint. In the structure, the ultra-long optical feedback waveguide substantially limits the scalable photonic RC integrated designs. In this paper, a time-delayed RC is proposed by utilizing a silicon-based nonlinear MRR in conjunction with an array of linear MRRs. These linear MRRs possess a high quality factor, providing sufficient memory capacity for the entire system. We quantitatively analyze and assess the proposed RC structure's performance on three classical tasks with diverse memory requirements, i.e., the Narma 10, Mackey-Glass, and Santa Fe chaotic timeseries prediction tasks. The proposed system exhibits comparable performance to the MRR with an ultra-long optical feedback waveguide-based system when it comes to handling the Narma 10 task, which requires a significant memory capacity. Nevertheless, the overall length of these linear MRRs is significantly smaller, by three orders of magnitude, compared to the ultra-long feedback waveguide in the MRR with optical feedback waveguide-based system. The compactness of this structure has significant implications for the scalability and seamless integration of photonic RC.

## 1. Introduction

Machine learning (ML) is a method that leverages data to increase performance in addressing challenging issues that are always beyond the capabilities of humans [1]. The use of machine learning (ML) has exploded over the past two decades in various areas, including recommendation systems, autonomous driving, image, speech, and text processing. A class of ML algorithms known as artificial neural networks (ANNs) is based on how the human brain functions. Recurrent neural networks (RNNs) are a special kind of artificial neural networks (ANNs) that are used to handle time-dependent input [2]. However, training RNN is notoriously challenging due to the vanishing and the exploding gradient problems. The objective function of a RNN model with many hyper-parameters is extremely time-consuming to optimize [2,3]. Reservoir computing (RC) is a computational framework that enables high-speed machine learning originated from RNN models. It can balance training complexity and performance, and has recently attracted the attention of many researchers [4-6]. In RC, a set of input signals are mapped into higher dimensional computational spaces using a nonlinear dynamical system called a reservoir. By exploring temporal correlations in the data, RC performs better than traditional feedforward neural networks. Only the nodes in the readout layer are trained, while the reservoir is made up of a network of thousands of nodes sparsely connected by fixed-random weights. As a result, the entire training process is linear, allowing RC to maintain good performance with low complexity.

However, in conventional RC systems, reservoir nonlinearity is frequently given by a sizable number of nonlinear nodes, which makes hardware implementation extremely challenging and results in complicated systems [7-8]. To solve this problem, RC has extended to physical systems that are continuous systems in space and/or time rather than networks in the traditional sense [7-8]. Photonic RC has the advantages of ultra-high operating bandwidth, low power consumption, parallel computing via signal multiplexing, which gives an ideal foundation for hardware acceleration [9-12]. Overall, spatial and time-delayed node reservoirs are the classifications given to photonic RC systems [13]. The structure of the former is comparable to that of an RNN whose nodes are spatially distributed and interconnected. Therefore, there are numerous photonic hardware requirements [14,15]. The time-delayed node reservoir, comprises of just one computing-related physical node and a number of time-multiplexed virtual nodes connected to an internal time-delayed feedback loop [16-20]. To implement RC, nonlinear components like photodetectors, semiconductor lasers, and electro-optic modulators have been incorporated into an optoelectronic delayed feedback loop [21-27]. These systems operate at GHz velocity and can support thousands of virtual nodes. Various disciplines, including chaotic time series prediction, image and speech recognition, nonlinear channel equalization, and chromatic dispersion compensation in optical communication systems [14,28-31], have seen considerable success because of the application of photonic RC.

In order to attain good performance in the near future, time-delayed node reservoirs must exponentially increase the number of virtual nodes and the complexity of their connectivity. This is due to the increasing demand for processing capabilities. However, scaling these parameters based on optoelectronic splitting systems becomes increasingly impractical for these time-delayed node architectures [32,33]. As a result, the integrated photonic RC designs are the subject of the current research. MRR is one of fundamental building blocks in integrated photonic devices, and has been employed to construct a time-delayed node reservoir because it exhibits a variety of rich nonlinear dynamical features [34,35]. Using silicon-on-insulator (SOI) MRRs as nonlinear nodes, a 4×4 swirl reservoir topology has been theoretically established to perform a traditional nonlinear Boolean problem in Ref. [36]. Through the use of its own linear and nonlinear memory capacities as well as virtual nodes created by time multiplexing, a single silicon-based MRR has also demonstrated its capacity to carry out tasks that need memory [37]. To create time-delayed reservoir computers, a single silicon-based MRR with an external optical feedback waveguide has recently been proposed [38]. The addition of an optical feedback waveguide allows for a large increase in the system's linear memory capacity (MC). To achieve a good memory capacity, the length of the feedback waveguide in the optical feedback system is optimized at about 20 cm. In the presence of optical feedback waveguide, the system is superior to a single silicon-based MRR for the task requiring huge amounts of memory. Nevertheless, this waveguide length is far longer than the microring waveguide's diameter. It is quite difficult to provide scalability and integrate technology into photonic RC systems with such a huge scale.

In this article, we present a silicon-based main cavity coupled in series by a number of linear cavity array to build a time-delayed RC with large MC. In this approach, the reservoir is emulated by using a single main cavity as a physical nonlinear node. This time multiplexing results in a distributed set of virtual nodes along the delayed feedback loop for the single main cavity response. In this study, the system's MC is greatly improved by means of an array of series-coupled linear cavities. We numerically analyze and assess the proposed RC's performance of on three classical tasks that need different compromises between nonlinearity and memory. When the task requires more memory than a single MRR can give, the proposed system dramatically improves computing performance. The article is structured as follows. Section 2 describes the model of the main cavity coupled in series by a number of linear cavities. In section 3, we describe its implementation in a time-delayed RC scheme. In section 4, we examine and explain the numerical results obtained for the Narma 10, Mackey-Glass, and Santa Fe chaotic timeseries prediction challenges. Finally, in section 5, we conducted a tolerance analysis of the proposed single-SCMRRs-10 configuration's performance in the NARMA 10 task. While the issue of resonant wavelength drift is inevitable, current technologies demonstrate a certain degree of adaptability to tolerate specific stochastic errors. It is noteworthy that viable solutions do exist for addressing the resonant wavelength drift arising from manufacturing imperfections.

## 2. Series coupled microring resonators

Figure 1 illustrates two configurations of series-coupled microring resonators (SCMRRs) for building time-delayed reservoir computers. As an all pass filter structure, both configurations consist of a waveguide and series-coupled MRRs side coupled to the waveguide. Among these series-coupled MRRs, only the MRR that directly coupled with the waveguide exhibits nonlinear behavior, while the others manifest only linear behavior. The former is referred to as the main cavity, and the latter are called the linear cavity array in this paper. In the configuration of Fig. 1(a), only one array of series-coupled MRRs is coupled to the main cavity, which is referred to as the single-SCMMRs. The structure is excited by the power signal $E_{in}$ at the entrance to the input waveguide. By leveraging nonlinear coupled mode theory [38], we describe separately the time evolution of the light wave amplitude, free carrier density, and temperature within the main cavity, as follows,

$$\frac{dU(t)}{dt} = [i(\omega_0(t) - \omega_p) - \gamma(t)]U(t) + i\mu E_{in}(t) + i\mu_1 U_1(t) \quad (1)$$

$$\frac{d\Delta N(t)}{dt} = -\frac{\Delta N(t)}{\tau_{FC}} + G_{TPA}|U(t)|^4 \quad (2)$$

$$\frac{d\Delta T(t)}{dt} = -\frac{\Delta T(t)}{\tau_{TH}} + \frac{P_{abs}}{M_{ring}c_{Si}}. \quad (3)$$

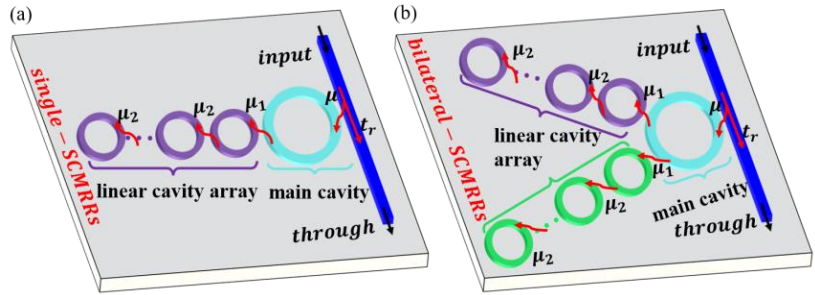

**Fig. 1.** Schematic diagram of two types of SCMRRs designed for constructing time delayed reservoir computers. (a) Schematic diagram of single SCMMRs, where only one array of series-coupled MRRs is coupled to the main cavity. (b) Schematic diagram of bilateral SCMMRs. The structure is composed of two arrays of series-coupled linear cavity arrays, which are interconnected with the main cavity. Each array of linear cavity array has the same resonance wavelength, but the two arrays of linear cavity arrays exhibit a difference in resonance wavelength.

Here, in Eq. (1), $U(t)$ represents the light wave amplitude in the main cavity, $U_1(t)$ is the light wave amplitude in the linear MRR adjacent to the main cavity, $\gamma(t)$ is the total loss rate, and $\mu$, $\mu_1$ are the coupling coefficients between the main cavity and the straight waveguide or the linear MRR of its nearest neighbor, respectively. Two coupling coefficients satisfy that $\mu = \sqrt{\gamma_{e1}}$ and $\mu_1 = \sqrt{\gamma_{e2}/\Gamma_{t1}}$, where $\gamma_{e1}$, $\gamma_{e2}$ represent separately the main cavity's extrinsic loss rates due to coupling to the waveguide and the linear MRR of its nearest neighbor, and $\Gamma_{t1}$ is the time it takes for the light to travel one round-trip in the main cavity. The generation of free carriers in the main cavity is attributed to two-photon absorption (TPA). And its production rate is indicated by $G_{TPA}$, while its recombination lifetime is expressed by $\tau_{FC}$. Both the production and the recombination of free carriers are influenced by the phonon emission in the silicon-based waveguide, leading to a consequent change ($\Delta T$) in its mode-averaged temperature. Eq. (3) is derived from Newton's law, where $\tau_{TH}$ represents the thermal decay time due to the heat dissipation in the surrounding medium, $P_{abs}$ is the absorption power of the material that is heated, $M_{ring}$ is the mass of the main cavity, and $c_{Si}$ is the heat capacity of the silicon-based waveguide.

The refractive index of the main cavity is adjusted by thermal and free carrier variations. This further causes the changes of the parameters $\omega_0(t)$ and $\gamma(t)$ in Eq.(1). The first term $\omega_0(t)$ is expressed as $\omega_0(t) = \omega_0 + \delta\omega_{nl}(t)$, where $\omega_0$ is the resonance frequency in the absence of nonlinearity, and $\delta\omega_{nl}(t)$ is the resonance frequency shift due to nonlinear contribution. In the systems, the nonlinear influence $\delta\omega_{nl}(t)$ can be written as,

$$\delta\omega_{nl}(t) = -\frac{\Gamma_c \omega_0}{n_{Si}}\left(\frac{dn_{Si}}{dT}\Delta T(t) + \frac{dn_{Si}}{dN}\Delta N(t)\right), \tag{4}$$

where $\Gamma_c$ represents the modal confinement factor and $n_{Si}$ is the refractive index of silicon. The second term $\gamma(t)$ contains the loss rate $\gamma_l$ in the linear condition and the loss rates due to TPA and free carrier absorption (FCA):

$$\gamma(t) = \gamma_l + \eta_{FCA}\Delta N(t) + \eta_{TPA}|U(t)|^2. \tag{5}$$

In Eq. (5), $\gamma_l = \gamma_{i1} + \gamma_{e1}$, $\gamma_{i1}$ represents the intrinsic loss rate of the main cavity, and $\eta_{FCA}, \eta_{TPA}$ represent separately the efficiency of FCA and TPA. When the main cavity is operated in a linear state, we can get the conditions of $\delta\omega_{nl}(t) = 0$ and $\gamma(t) = \gamma_l$. Then, the characteristic time-scale of the main cavity in absence of nonlinearity is decided by its photon lifetime $\tau_{ph} = \gamma_l^{-1}$. As the nonlinear state is induced by the TPA, there are two important timescale parameters ($\tau_{FC}$ and $\tau_{TH}$) present for the dynamic evolution of the system. $\tau_{FC}$ is about two orders of magnitude smaller than $\tau_{TH}$. When the timescale of the input signal is coincident with a timescale parameter, only corresponding nonlinear effects exert influence on the dynamics. In the paper, the input signal is encoded at the timescale of $\tau_{FC}$, and we emphasize the nonlinear effect activated by the free carriers in the main cavity.

In the SCMRRs, the time evolution of light wave amplitudes in other linear cavity array is derived by the following set of coupled differential equations:

$$\frac{dU_1(t)}{dt} = i(\omega_1 - \omega_p)U_1(t) + i\mu_1 U(t) + i\mu_2 U_2(t) \tag{6}$$

$$\frac{dU_M(t)}{dt} = i(\omega_1 - \omega_p)U_M(t) + i\mu_2 U_{M-1}(t) \tag{7}$$

$$\frac{dU_m(t)}{dt} = i(\omega_1 - \omega_p)U_m(t) + i\mu_2 U_{m-1}(t) + i\mu_2 U_{m+1}(t), \tag{8}$$

where these series coupled linear cavities are numbered as the indices $m$ from 1 to $M$ ($M$ is the total number of these series-coupled linear cavities). For simplicity, we assumed that all the linear cavities are identical, and the coupling coefficients between any two adjacent cavities are equal. Here, $\omega_1$ is the resonance frequency of the linear MRR, and $\mu_2 = \sqrt{\gamma_{e2}/\Gamma_{t2}}$ is the coupling coefficient between two adjacent linear cavities, where $\Gamma_{t2}$ is the time it takes for the light to travel one round-trip in the linear MRR. We designated the index of the linear MRR adjacent to the main cavity as $m=1$, and the variation of its light wave amplitude is described by Eq. (6). Eq. (7) describes the optical energy amplitude variation within the last linear MRR at $m=M$. For the $m^{th}$ ($m = 2, L, M-1$) series coupled MRR located between the first and last linear cavities, the evolution of its light wave amplitude is given by Eq. (8).

The output signal $E_{th}$ through the waveguide can be expressed as,

$$E_{th}(t) = t_r E_{in}(t) + \mu U(t), \tag{9}$$

where $t_r$ represents the field transmission from the input port to the through port.

Figure 1(b) displays a SCMRRs-based structure that resembles the configurations shown in Fig.1 (a). This structure consists of two arrays of series-coupled linear cavity arrays, each coupled to the main cavity. Within each array of linear cavity array, all linear cavities in each linear cavity array share the same structural and physical parameters, and the coupling coefficients between any two adjacent linear cavities are equal. Consequently, all linear cavities in each array possess the same diameter and resonance wavelength. However, there is a difference in the resonance wavelength between the two arrays of linear cavity arrays. In the paper, this structure is referred to as bilateral SCMMRs. The detailed coupled mode equations for this configuration are provided insupplement. To simplify the nomenclature, the system with varying numbers of series-coupled linear cavities is denoted using an abbreviated version. For instance, if the single SCMRRs system has 2 series-coupled linear cavities, it is referred to as single SCMRRs-2; Similarly, if the bilateral SCMRRs system has two arrays of 2 series-coupled linear cavities, it is named bilateral SCMRRs-2.

The initial wavelength detuning between the laser wavelength and the main cavity resonance is defined as $\Delta\lambda_s = \lambda_p - \lambda_0$. The resonance shift caused by nonlinear effects is denoted by $\Delta\lambda_0(t) = \lambda_0(t) - \lambda_0$, where $\lambda_0 = 2\pi c/\omega_0$ and $\lambda_0(t) = 2\pi c/\omega_0(t)$. The set of coupled differential Eqs. (1)-(8) is numerically solved using the Runge-Kutta method with an integration time step of 2ps, which is significantly smaller than the lowest timescale effects ($\tau_{ph} \approx 97\text{ps}$). Before problem-solving, these equations are transformed into dimensionless form for convenience (see supplementfor details). In this model, only one-way propagation is considered, and the values of the parameters are provided in supplementwhich includes a detailed table with parameter values.

## 3. SCMRRs in time-delayed RC

Figure 2 depicts the schematic of the proposed time-delayed RC system, which comprises an input layer, a reservoir, and an output layer [7]. In the input layer, the time-continuous input signals are first encoded as a sequence of bits, where $x_i$ represents the amplitude of the $i$th bit, and $\tau$ represents the bit period. Subsequently, a bit mask is applied by multiplying the bit stream with a set of random values $M(t)$. $M(t)$ is a periodic sequence with a period $\tau$, and it follows uniform distribution. The resulting signal is then modulated onto the intensity of the optical carrier. The maximum laser input power is indicated by $P_M$. Next, the modulated optical signal enters the single SCMRRs through the entrance to the input port and propagates within all the MRRs' waveguide through the coupling between the MRRs. The received optical signal at the through port is converted into an electrical signal by a photodetector (PD). Within one period of time duration $\tau$, the electrical signal is sampled synchronously at the masking sampling interval $\theta$, resulting in $N$ equidistant sampling points equally spaced in time by $\theta = \tau/N$. These $N$ equidistant points are defined as $N$ virtual nodes, and they play a role similar to one of the nodes in a conventional reservoir. In the reservoir layer, the system's nonlinear characteristics are generated by the main cavity and the PD, while the series coupled linear cavity array significantly enhance its MC. Consequently, the original input signal is nonlinearly transformed from the physical system into a higher dimensional space with $N$ virtual nodes. Ultimately, in the output layer, a predicted value $o_i$ corresponding to the input $x_i$ is determined by a linear combination of the responses of the related virtual nodes, as follows,

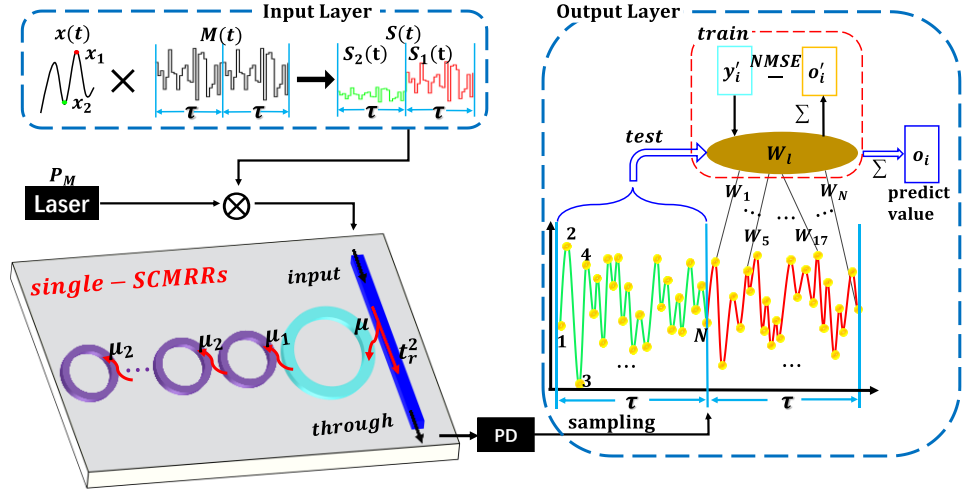

**Fig. 2.** Schematic of the time-delayed RC with single-SCMRRs. The input information $x(t)$ is first masked by a sequence $M(t)$. The masked signal is then modulated onto the intensity of the optical carrier excited by the laser. At the through port, the corresponding electrical signal is collected by the PD. The virtual nodes of the reservoir are created through time-multiplexing, and a predicted value $o_i$ is obtained by a linear weighted sum of the responses of the related virtual nodes. During the training process, the optimal output weights are obtained by utilizing the expected values $y_i$ derived from the original dataset.

$$o_i = \sum_{l=1}^{N} W_l N_{l,i} \qquad (10)$$

where $N_{l,i}$ is an element value of an $N$-dimensional vector, which is the response of the virtual nodes at the $i$th period, and $W_l$ is its corresponding readout weight. The weights of the readout layer are trained by utilizing a ridge regression method to minimize the normalized mean square error (NMSE) between $o_i^{'}$ and the expected value $y_i^{'}$, which is expressed as,

$$NMSE = \frac{\langle \| o_i^{'} - y_i^{'} \|^2 \rangle}{\langle \| y_i^{'} - \langle y_i^{'} \rangle \|^2 \rangle} \qquad (11)$$

The main cavity has a quality factor of $1.18\times10^5$, and self-pulsations can occur dependent on a free carrier concentration variation in the microring waveguide. The system's MC can be significantly enhanced by introducing the series coupled linear cavity array. The masking sampling interval is set to $\theta=40\text{ps}$ [38], which is considerably lower than three different timescales: the photon lifetime (here $\tau_{ph}\approx 97\,ps$), the thermal lifetime (here $\tau_{TH}\approx 83.3\text{ns}$) and the free carrier lifetime (here $\tau_{FC}\approx 3\text{ns}$) in the main cavity. Consequently, the system operates in a transient state, and the current internal field in the main cavity does not fully dissipate when the next mask signal arrives. As a result, the main cavity exhibits a short memory, and adjacent virtual nodes are coupled via inertia. There is no doubting the fact that the presence of the series-coupled linear cavity array strengthens inertia. The free carrier nonlinearity has a faster time response, and to achieve GHz computing rates, the bit period is set to $\tau=1ns$. In this case, the number of virtual nodes is chosen as $N=25$ to ensure compatibility with delays generated by series-coupled linear cavity array. When the masked optical signal enters the main cavity, free carriers are generated, leading to a resonance shift $\Delta\lambda_{FC}$. Simultaneously, the resonance is also shifted by $\Delta\lambda_{TH}$ due to thermal effects. However, because the signal speed is much faster than these variations, the nonlinear transformation of the optical signal does not occur, and only resonance shifts are induced. The main cavity with high optical power always exhibit self-pulsing dynamics, and thermal effects contribute to the nonlinear transformation. On one hand, when the onbit period $\tau$ is smaller than $1ns$, the number of virtual nodes is restricted, resulting in a decline in the computational performance of the reservoir. On the other hand, a larger number of virtual nodes can be obtained at $\tau>1ns$ and the computational performance can be significantly improved, but at the cost of computing speed. Thus, the bit period is selected as $\tau=1ns$ to strike a balance between computational speed and performance.

## 4. Results

Using the numerical methods described earlier, three classical computational tasks named NARMA 10, Mackey-Glass, and Santa Fe, are employed to evaluate the computational performance of the proposed RC systems based on single and bilateral SCMRRs. The NARMA 10 task is a discrete-time 10*th* nonlinear autoregressive moving average (NARMA) system [39]. It is widely employed for testing Echo State Networks (ESN) models, which are a type of RC utilizing a RNN with a sparsely connected hidden layer [40]. The NARMA 10 task requires considering at least the previous 10 values to predict the next value, demanding a large amount of MC. The task involves both nonlinear transformation and memory. The Mackey-Glass time series serves as a standard benchmark for chaotic time series prediction tasks [41]. The Santa Fe laser time series involves one-step-ahead prediction on data acquired by sampling the intensity of a far-infrared laser in a chaotic state [42]. For both the Mackey-Glass and Santa Fe tasks, the next value $x_{i+1}$ is solely related to the current value $x_i$, requiring a short MC. These tasks have varying demands for signal processing and MC, making their computing performances suitable for evaluating the overall efficiency of a neuromorphic computing system.

According to RC theory, nonlinear dynamics plays a crucial role, but they might degrade MC. To achieve better computational performance, the RC system needs to strike a balance between the nonlinear transformation of the input information and MC. However, it is challenging to determine the extent of nonlinear transformation and the amount of MC required for a specific task. Therefore, it is essential to estimate or adjust the degree of nonlinearity and memory separately. The degree of nonlinearity can be evaluated indirectly by the standard deviation of resonance wavelength shift $\sigma(\Delta\lambda_0(t))$ in the main cavity. A larger standard deviation indicates stronger nonlinearity, and vice versa. The MC mainly refers to the linear MC, which is one of the fundamental requirements of RC, it can be calculated by training the reservoir to reconstruct an input stream of values [0, 0.5] with a uniform distribution $k$ timesteps later [43]. The MC can be given by:

$$MC = \sum_{k=1}^{l_{max}} MC_k, \tag{12}$$

$$MC_k = \frac{\text{cov}^2(x_{i-k}, y_k)}{\text{var}(x_i)\text{var}(y_k)} = 1 - NMSE, \tag{13}$$

where $MC_k[0,1]$ is the MC for a $k$-bit shift, which represents the theoretical upper limit for the summation in the same work. On one hand, $MC_k = 1$ reflects a perfect memory of the bit stream $k$ bits later. On the other hand, $MC_k = 0$ means that all memory is lost, indicating no capability to recall past information. In this context, $l_{max}$ represents the calculated maximum length of memory sequences, and $\text{var}(\cdot)$, $\text{cov}^2(\cdot)$ denote the variance of a random variable and the covariance between two vectors, respectively.

Subsequently, we investigate the influence of the critical operational parameters on the performance of the three selected tasks based on the SCMRRs system, including the maximum input laser power $P_M$, the initial wavelength detuning $\Delta\lambda_s$, the ratio ($Q_2/Q_1$) of the linear MRR's quality factor to the main cavity's quality factor, and the total number $M$ of the series-coupled linear cavities in each array. The main cavity has a radius of 6.75 $\mu$m and a quality factor of $1.18\times10^5$, while all the linear cavities have a radius of 1.51 $\mu$m. These parameters have a significant impact on the nonlinear dynamics in the main cavity, and the number and quality factor of the series-coupled linear cavities greatly affect the system's MC. In the paper, the maximum input laser power $P_M$ is varied from 0.1mW to 7mW, the initial wavelength detuning $\Delta\lambda_s$ is adjusted from -30pm to 30pm with a step size of 5pm to cover all the main cavity resonance (full width at half maximum satisfies that $FWHM$=26pm), the radio ($Q_2/Q_1$) of the linear MRR's quality factor to the main cavity's quality factor is adjusted from 10 to 500. Additionally, the maximum total number of these linear cavities is set to 10. For each parameter variation, previous 1000 data are firstly input into the system to remove any fluctuations induced by the inclusion of inputs. Then, 2000 input data are used for the training, and the next 1000 data are used for the test data. The same data was not shared between the training set and the test set. All the simulations employ the same random mask, and the total number of virtual nodes is 25 by default in the article ($\tau = 1$ns). The linear classifier in RC's output layer involves the operation of the ridge regression, and the ridge regression coefficient is set to $10^{-4}$.

### 4.1 NARMR 10 benchmark test

The output of the NARMA 10 system is described as follows:

$$y_{i+1} = 0.3y_i + 0.05y_i\sum_{k=0}^{9}y_{i-k} + 1.5x_{i-9}x_i + 0.1, \tag{14}$$

where $x_i$ is a random input at the $i$th moment, generated from a uniform distribution within the range [0, 0.5], and $y_i$ is the corresponding output at the $i$th moment. The readout network is trained to predict $y_i$ from the reservoir state and $x_i$. The task requires predicting the next output value based on at least 10 output values (the current one and the previous 9 values), indicating a significant need for memory. As previously mentioned, the computing performance depends mainly on four critical parameters: the maximum input laser power $P_M$, the initial wavelength detuning $\Delta\lambda_s$, the ratio ($Q_2/Q_1$) of the linear MRR's quality factor to the main cavity's quality factor, and the total number $M$ of the series-coupled linear cavities. The first two parameters determine the nonlinear strength of the main cavity, while the latter two parameters are related to the MC of the main cavity. The NARMA 10 task requires a large MC, and does not require strong nonlinear signal transformation. For a given quality factor ratio ($Q_2/Q_1$) and the total number (*M*) of the linear cavities in each array, optimal performance occurs at the maximum input laser power $P_M = 0.1\text{mW}$ and the initial wavelength detuning $\Delta\lambda_s = -10\text{pm}$. The injected laser power and detuning use the same values as those in Ref. [38], which are adopted for the remainder of Section 4.1. In this case, the main cavity operates in the linear state.

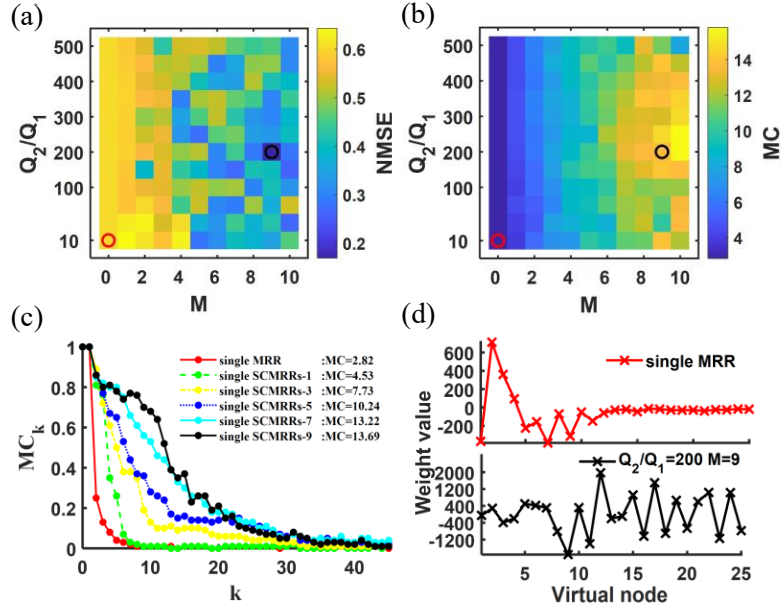

**Fig. 3.** Performance of the NARMR 10 benchmark task for the proposed single SCMRRs-based RC. (a) NMSE and (b) MC versus the quality factor's ratio $Q_2/Q_1$ and the total number *M* of the linear cavities for the single SCMRRs system. (c) MC (memory function $MC_k$, with $l_{max}=45$) of a single SCMRRs-based system under different numbers of linear MRR. (d) The calculated weight values for the task to remember the previous input value $x_{i-1}$ based on the single main cavity (red curve) and the proposed SCMRRs system that result in the lowest NMSE (black curve).

Figure 3 illustrates the performance of the NARMA 10 benchmark task for the single SCMRRs system. Fig. 3 (a) and (b) display separately NMSE and MC versus the ratio ($Q_2/Q_1$) of their quality factors and the total number *M* of the linear cavities for the single SCMRRs system. The NMSE achieves its lowest value when the MC reaches its maximum value. With the initial increase of the linear MRR's number or quality factor, the stronger the feedback strength of the system is, and the longer the photonic lifetime becomes. This results in an extended linear memory. The minimum error $NMSE_{min}$=0.169 is found at the quality factor's ratio $Q_2/Q_1 = 200$ and the total number of the linear cavities $M = 9$ for $P_M = 0.1\text{mW}$ and an initial wavelength detuning of $\Delta\lambda_s = -10\text{pm}$. Specifically, the main cavity's resonance wavelength is 1549.66nm in the absence of nonlinearity, while the linear MRR's resonance wavelength is 1549.71nm. When the linear MRR's number or quality factor continues to increase from the position of minimum *NMSE* value, the power coupled into the latter series-coupled MRRs decreases. Therefore, even if the number of series-coupled MRRs exceeds 9, the actual number of series-coupled MRRs is limited to 9, where the corresponding MC reaches its maximum value.

Fig. 3 (c) shows the comparison of memory function $MC_k$ between the single MRR-based RC and the proposed SCMRRs-based RC. The single main cavity has a limited amount of memory, which arises from the inertia between the responses of the first virtual nodes to the current input value $x_i$ and the responses of the last virtual nodes to the last input value $x_{i-1}$. The single MRR-based RC can only remember the last input value $x_{i-1}$ from the current input $x_i$ (Fig. 3 (c), red curve). Fig. 3 (d) displays the computed readout weights for the task to remember the previous input value $x_{i-1}$ (red curve). Since the response of the reservoir to the actual input $x_i$ is considered during the training step, the calculated results indicate that only the weight values of the first several virtual nodes mainly contribute to the computation due to the limited MC. In contrast, in the single SCMRRs system, these series-coupled linear cavities act as a linear analog shift register. As depicted in Fig. 3(c), their memory storage capacities are significantly enhanced with an increase in the total number of series-coupled linear cavities in each array. The main cavity is initially excited by the optical signal injected from the input waveguide. When the optical signal's frequency is close to the resonant frequency of these series-coupled linear cavities, a portion of the optical signal is coupled gradually from the main cavity to these linear cavities. These signals propagate many round-trips through these linear cavities and are eventually coupled back to the main cavity. Due to optical signals are continuously coupled into these series-coupled linear cavities with a high quality factor, the proposed SCMRRs-based RC obtains an extended linear memory. As shown in Fig. 3 (d) (black curve), almost all virtual nodes contribute to the task computation, indicating a significant improvement in MC compared to the single MRR-based RC.

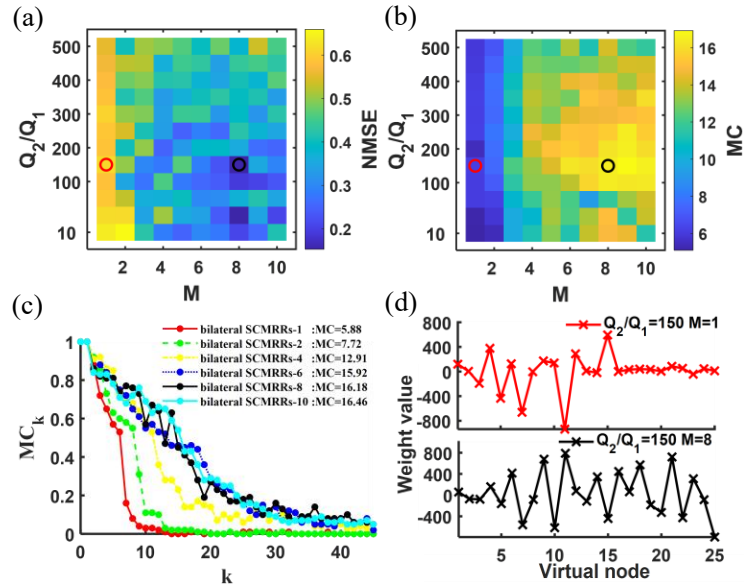

**Fig. 4.** Performance of the NARMA 10 benchmark task for the proposed bilateral SCMRRs based RC. (a) NMSE and (b) MC versus the quality factor's ratio $Q_2/Q_1$ and the total number $M$ of the linear cavities in each array for the bilateral SCMRRs based system. (c) MC (memory function $MC_k$, with $l_{max}=45$) of a bilateral SCMRRs based system under different numbers of linear cavities in each array. (d) The calculated weight values for the task to remember the previous input value $x_{i-1}$ based on the proposed bilateral SCMRRs-1 system (red curve) and bilateral SCMRRs-8 system that result in the lowest NMSE (black curve).

Figure 4 displays the performance of the NARMA 10 benchmark task for the bilateral SCMRRs system. Fig. 4 (a) and (b) display separately NMSE and MC versus the ratio ($Q_2/Q_1$) of their quality factors and the total number $M$ of the linear cavities in each array for the bilateral SCMRRs system. The NMSE achieves its lowest value when the $MC$ is close to its maximum value for the bilateral SCMRRs system. In this case, the bilateral SCMRRs system works in a linear state at the maximum input laser power $P_M=0.1\text{mW}$ and the initial wavelength detuning

$\Delta\lambda_s = -20\text{pm}$. Specifically, the main cavity's resonance wavelength is 1549.66nm in the absence of nonlinearity, and the resonance wavelengths of two arrays of the linear cavity array are 1549.71nm and 1549.74nm, respectively. Fig. 4 (c) shows the memory function $MC_k$ of the proposed bilateral SCMRRs-based RC. At the quality factor's ratio $Q_2/Q_1=150$ and the total number of the linear cavities in each array $M=1$, the MC is 5.88, which is close to the minimum MC value. The MC corresponds to the maximum NMSE of 0.65. The minimum NMSE of 0.154 appears at $Q_2/Q_1=150$ and $M=8$. The corresponding MC is 16.18, which is close to the maximum MC value of 16.46. In the former, due to lack of memory, only the first virtual nodes have nonzero weight values, which are important for computation (red curve, Fig. 4 (d)). In the latter, with sufficient MC, the system has non-zero weights at almost all virtual nodes (black curve, Fig. 4 (d)). As a result, the responses of almost all virtual nodes contribute to the task computation, leading to a low prediction error. In contrast to the single SCMRRs system, the bilateral SCMRRs system possesses two arrays of series-coupled linear cavity arrays with different resonance wavelengths. Thus, its MC can be increased in the wavelength dimensions compared with the corresponding single SCMRRs.

The performance of the proposed SCMRRs based RC is evaluated against several MRR-based RCs, including the single MRR without optical feedback and the MRR structure with external optical feedback. For the sake of fairness, these comparisons are made under the same environmental conditions, including the MRR's structural and material parameters and the number of virtual nodes in the reservoir. The number of virtual node has a huge impact on the NMSE of the photonic RC, with a larger number of virtual nodes leading to more accurate results. Therefore, in this paper, the number of virtual nodes is fixed at 25, utilizing the masking sampling interval $\theta = 40\text{ps}$ and the bit period $\tau = 1\text{ns}$. Table 1 shows the NMSE comparison of the proposed SCMRRs based RC and several MRR-based RCs for the NARMA 10 task. At $P_M = 0.1\text{mW}$ and $\Delta\lambda_s = -20\text{pm}$, the bilateral-SCMRRs-8 system achieves the lowest prediction error (NMSE=0.154).

Table 1. The NMSE comparison of the proposed SCMRRs-based RC and several MRR-based RCs for the NARMA 10 task.

| | single MRR | MRR with feedback | single-SCMRRs-9 | bilateral-SCMRRs-8 |
|---|---|---|---|---|
| Model | 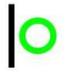 | 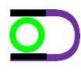 | 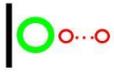 | 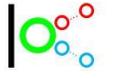 |
| NMSE | 0.481 | 0.187 | 0.169 | 0.154 |

For the Narma 10 task, both memory and nonlinearity contribute to the task computation. In these photonic RC systems, nonlinearity derives from both main cavity nonlinearity and PD nonlinearity. Because of the small input laser power ($P_M = 0.1\text{mW}$), the main cavity operates in the linear state, making the PD nonlinearity the dominant factor [38]. Large memory is a key factor in improving computational accuracy. Both the proposed SCMRRs based system and the MRR with optical feedback can provide sufficient memory, resulting in lower NMSE than others. However, the MRR with optical feedback has a notable drawback, as the length of its feedback waveguide spans approximately 20 centimeters. Such a long waveguide faces many substantial challenges in applications, including device fabrication, transmission loss, temperature control, etc. The proposed SCMRRs system greatly enhances the MC by employing multiple series-coupled MRRs, and shares the same computational performance with the MRR with optical feedback. At the same time, the proposed SCMRRs based system has an ultra-small size due to the use of resonant feedback structures. Furthermore, recent scientific breakthroughs have been made in the preparation of series-coupled MRRs [44-45], making it feasible to fabricate the proposed SCMRRs using existing fabrication techniques.

## 4.2 Mackey-Glass benchmark test

The Mackey-Glass chaotic time series is defined by the following differential equation:

$$\frac{dy(t)}{dt} = \frac{0.2y(t-\tau)}{1+y(t-\tau)^{10}} - 0.1y(t) \quad (15)$$

where $y(t)$ is the output at time step $t$, and $\tau$ is the time delay. The reservoir computing task is to predict the value $\delta$ steps ahead for a time series stemming from a Mackey-Glass delay equation (Eq. 15) with $\tau = 17$. We solve Eq. (15) numerically by using the fourth-order Runge-Kutta method with an integration step of 0.1 to exhibit moderate chaotic dynamics [38]. After solving the differential equation, we obtain a continuous time series. To perform the RC task, the continuous time series is downsampled with a fixed time interval of $t_s=3$ to obtain a discrete time series $y_k$. This discrete-time series is then used for the prediction task. The dataset of 3000 values was separated into 2000 samples for training and 1000 samples for testing.

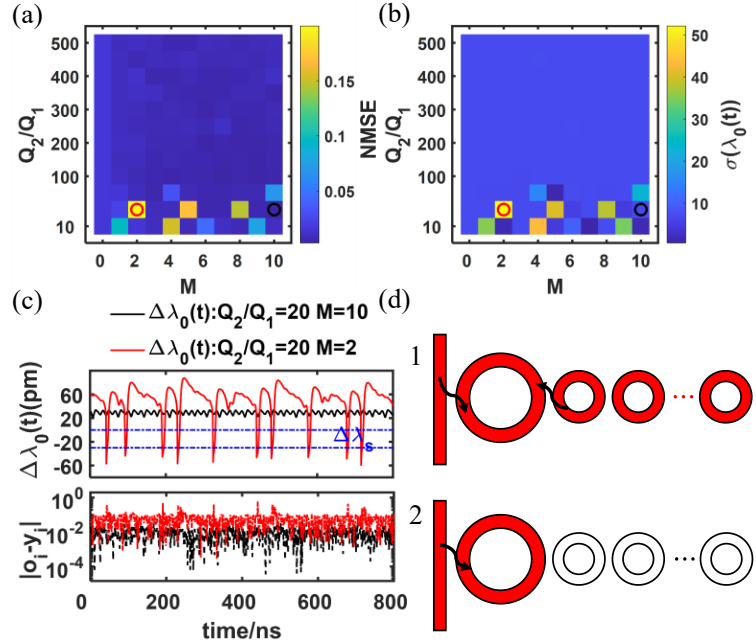

**Fig. 5.** Performance of the Mackey-Glass benchmark task for the proposed SCMRRs based RC. (a) NMSE and (b) $\sigma(\Delta\lambda_0(t))$ versus the quality factor's ratio $Q_2/Q_1$ and the total number M of these linear cavities for the single SCMRRs based system. (c) The resonance shift and the bit error versus time during the task have two types of extreme NMSE values (maximum and minimum): the black curve corresponds to the lowest NMSE (black circle in (a)) and the red curve corresponds to the largest NMSE (red circle in (a)). (d) Dynamical evolution of the single-SCMRRs based system exhibiting the self-pulsation phenomenon: light is coupled into (path 1, upper) these linear cavities or does not enter (path 2, lower) these cavities.

Figure 5 displays the performance of the Mackey-Glass benchmark task for the single SCMRRs based system. As shown in Fig. 5 (a), the lowest NMSE of the predicted value occurs at the quality factor's ratio $Q_2/Q_1=20$ and the total number of these linear cavities $M=10$ for an input laser power of $P_M = 6\text{mW}$ and an initial wavelength detuning of $\Delta\lambda_s = -30\text{pm}$. On the whole, system nonlinearity plays a key role in the Mackey-Glass benchmark test, which is quite distinct from the NARMA 10 task. The system nonlinear response of the system is proportional to the intracavity, which is assessed by the standard deviation of the main cavity's resonance shift $\sigma(\Delta\lambda_0(t))$. Obviously, the initial resonance shift should be limited to a certain range so that the input power can be injected to the resonance of the main cavity.

The Mackey-Glass task does not require too much MC, and the single MRR without optical feedback achieves relatively good performance (NMSE=0.016) under the same conditions. In our system, the main cavity contributes to the nonlinearity, whereas the coupled linear cavity array serves as the memory provider. The memory is effective as long as the resonances of these cavities match each other. Meanwhile, the nonlinear response of the cavity shifts the resonance, which is also influenced by the SCMRRs. Therefore, there exists a trade-off between the nonlinearity and the MC. As shown in Fig. 5 (a), as the quality factor of the coupled linear cavity array is rather large, the single SCMRRs-based system shares similar NMSE performance with the single MRR without optical feedback. With the increase in these reservoir cavity array' quality factors, the narrow cavity resonance cannot fall into the broad resonance of the nonlinear cavity, resulting a limited MC. Compared with the single MRR without optical feedback, the single SCMRRs based system does not remarkably improve its MC. When the quality factor of the linear cavity array becomes small, the broad cavity resonance ensures significant coupling between the two MRRs, which modifies the optical power in the processing cavity and ultimately determines its nonlinearity. As shown in Figs. 5 (a) and (b), at $Q_2/Q_1=20$ and $M=2$, there is the highest NMSE, which corresponds to the largest $\sigma(\Delta\lambda_0(t))$. In this case, the series-coupled linear cavity array leads to high power in the main cavity, which further results in a large value of $\sigma(\Delta\lambda_0(t))$ and high system nonlinearity. The SCMRRs based system is in a seriously detuned state, and its NMSE performance is severely degraded. Fig. 5 (c) shows the resonance shift $\Delta\lambda_0(t)$ versus time when operating the computation. At $Q_2/Q_1=20$ and $M=2$, the main cavity's resonance wavelength shift versus time is indicated by the red curve in Fig. 5 (a). The multiple-peaked bursts appear in the curve. In this time period, the system generates too much detuning between the resonance wavelength and the input wavelength. The self-pulsation phenomenon takes place along with a thermal warming-up step and then a thermal cool-down step [46-47]. As shown in Fig. 5 (d) (path 2), the light signals mainly propagate through the main cavity during these time intervals, and they are not coupled into these series-coupled linear cavity array. Consequently, the proposed system has higher nonlinearity, but loses a lot of MC. In this way, the proposed system eventually produces a relatively large NMSE error. For comparison, we also find the lowest NMSE value at $Q_2/Q_1=20$ and $M=10$, which is indicated by the black circle in Fig. 5 (a). Its resonance wavelength shift versus time is indicated by the black curve in Fig. 5 (c). The value of $\Delta\lambda_0(t)$ is changed slightly with a small detuning with respect to the input wavelength. The self-pulsation phenomenon does not occur, and the light signal is coupled into series-coupled linear cavity array in the time interval (Fig. 5 (d), path 1).

Table 2. The NMSE comparison of the proposed SCMRRs-based RC and several MRR-based RCs for the Mackey-Glass task.

| Model | single MRR | MRR with feedback | single-SCMRRs-10 | bilateral-SCMRRs-8 |
|---|---|---|---|---|
| | 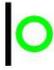 | 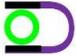 | 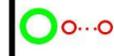 | 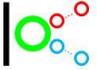 |
| NMSE | 0.016 | 0.0014 | 0.002 | 0.0018 |

Table 2 shows the NMSE comparison of the proposed SCMRRs-based RC and several MRR-based RCs for the Mackey-Glass task. The single SCMRRs-10 system achieves the minimum NMSE of 0.002, and the bilateral SCMRRs-8 system achieves the minimum NMSE of 0.0018. Both of them are far lower than the minimum NMSE of 0.016 obtained by the single MRR without optical feedback. The optimal NMSE is slightly larger than the result of the MRR with optical feedback (NMSE = 0.0014) [38]. The results show clearly that the Mackey-Glass task requires the combination of nonlinear transformation and MC for optimal NMSE performance. By combining the main cavity's nonlinearity with the memory provided by the series-coupled linear cavity array, the proposed system with small sizes achieves almost the same minimum NMSE with the MRR with optical feedback.

*4.3 Santa Fe benchmark test*

In the Santa Fe prediction task, the goal is to predict only the future step, and the memory provided by the single MRR without optical feedback is sufficient [38]. Additionally, this task also requires moderate system nonlinearity. Compared with the Narma 10 and the Mackey-Glass timeseries task, the available Santa Fe dataset contains experimental noise in its values, this noise introduces additional challenges in accurately predicting the future step.

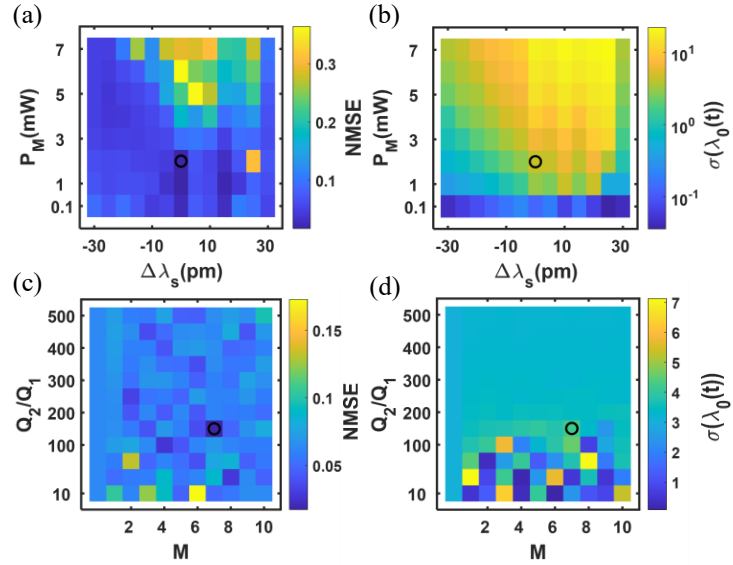

**Fig. 6.** Performance of the Santa Fe task based on the single SCMRRs system. (a) NMSE and (b) $\sigma(\Delta\lambda_0(t))$ versus the maximum input laser power $P_M$ and the total number $M$ of these linear cavities. (c) NMSE and (d) $\sigma(\Delta\lambda_0(t))$ versus the quality factor's ratio $Q_2/Q_1$ and the total number $M$ of these linear cavities.

Figure 6 displays the calculated performance of the Santa Fe task in the single SCMRRs based system. As shown in Fig. 6 (a), the minimum NMSE of 0.0173 is found at $P_M = 2$mW and $\Delta\lambda_s = 0$pm for $Q_2/Q_1$=150 and $M$=7, which is indicated by the black circle. Fig. 6 (b) shows the value of $\sigma(\Delta\lambda_0(t))$ versus the input laser power $P_M$ and the initial wavelength detuning $\Delta\lambda_s$ for the single SCMRRs based system. Correspondingly, $\sigma(\Delta\lambda_0(t))$ at the minimum NMSE shows a moderate value, which verifies that the system needs moderate nonlinearity for a low NMSE. As shown in Figs. 6 (c) and (d), we also separately discuss the influence of the quality factor's ratio $Q_2/Q_1$ and the total number $M$ of these linear cavities on the performance at $P_M = 2$mW and $\Delta\lambda_s = 0$pm. Under these conditions, the single SCMRRs system still achieves the minimum NMSE value at $Q_2/Q_1$=150 and $M$=7. And then the NMSE performance shows a slight change to the quality factor's ratio and the linear MRR's number. It is noted that $\Delta\lambda_0(t)$ oscillates with the change of the linear MRR's number at $Q_2/Q_1$<100, which is similar to the behavior of the single SCMMRs based system for the Mackey-Glass task under the same conditions. In Fig. 6 (d), $\sigma(\Delta\lambda_0(t))$ at the lowest NMSE also corresponds to an intermediate value.

Table 3. The NMSE comparison of the proposed SCMRRs-based RC and several MRR-based RCs for the Santa Fe task.

| Model | single MRR | MRR with feedback | single-SCMRRs-7 | bilateral-SCMRRs-5 |
|---|---|---|---|---|
| | 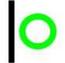 | 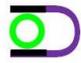 | 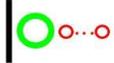 | 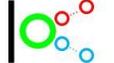 |
| NMSE | 0.031 | 0.020 | 0.017 | 0.018 |

Table 3 shows the NMSE comparison of the proposed SCMRRs based RC and several MRR-based RCs for the Santa Fe task. The single SCMRRs-7 system achieves the minimum NMSE of 0.017, which is very close to the minimum NMSE of 0.031 obtained by the single MRR without optical feedback. The optimal NMSE error of the single SCMRRs-7 is a little better than the result of the MRR with optical feedback (NMSE = 0.020) [38]. The results clearly demonstrate that the Santa Fe task only requires a moderate level of nonlinear transformation for achieving optimal performance. Since the task doesn't demand a significant amount of MC, only the MC of the single MRR without optical feedback is enough. Therefore, several MRR-based systems achieve almost the same optimal performance by employing their respective modest nonlinearities under different conditions. This indicates that for the Santa Fe task, the computational efficiency is mainly determined by the system's ability to handle moderate nonlinearity rather than extensive memory capabilities.

## 5. Tolerance analysis

In this study, an in-depth tolerance analysis of the proposed SCMRRs-10 configuration was conducted with respect to the NARMA 10 task. The variable Δr was defined as the radius error of the linear cavities. Ten distinct sets of linear cavities were meticulously selected by introducing random variations in Δr within the ranges of ±10 pm, ±50 pm, ±100 pm, and ±1 nm. The graphical representation in Fig. 7 vividly presents the MC and NMSE data under this specified setup.

The trends exhibited in the Fig. 7 are remarkably distinct and thought-provoking: the MC and NMSE data exhibit a pronounced stratified pattern in response to variations in Δr. As Δr fluctuates within the smaller range (±10 pm), the MC values are significantly maximized, accompanied by correspondingly minimized NMSE values. Conversely, when Δr fluctuates within the larger range (±1 nm), the MC values undergo substantial reduction, converging towards a single silicon-based MRR structure, resulting in notably inferior NMSE performance. This phenomenon originates from deviations in the fabrication process of the microcavities, inducing resonant wavelength drift in the linear cavities. Such drift adversely impacts the coupling efficiency of optical signals, consequently diminishing the system's memory capacity. In accordance with the citation of reference [51], investigators adroitly harnessed electron beam lithography and inductively coupled plasma dry etching methodologies to fabricate the indispensable active electrode silicon photonic high-order filters. By recourse to thermal tuning strategies, they efficaciously ameliorated the resonant wavelength perturbation conundrum. Moreover, they adroitly harnessed complementary metal-oxide-semiconductor (CMOS) fabrication techniques to instantiate a cascaded 10th-order adiabatic elliptical microresonator (AEM), thereby culminating in the fruition of a compact yet efficaciously performant silicon photonic filter [52]. The noteworthy confluence in the resonance wavelength across all microcavities is a salient observation, underscoring the extant potential of AEM technology to accommodate certain stochastic idiosyncrasies intrinsic to the fabrication process. By the same token, this vouchsafes the feasibility of redressing resonant wavelength perturbation issues engendered by manufacturing imperfections.

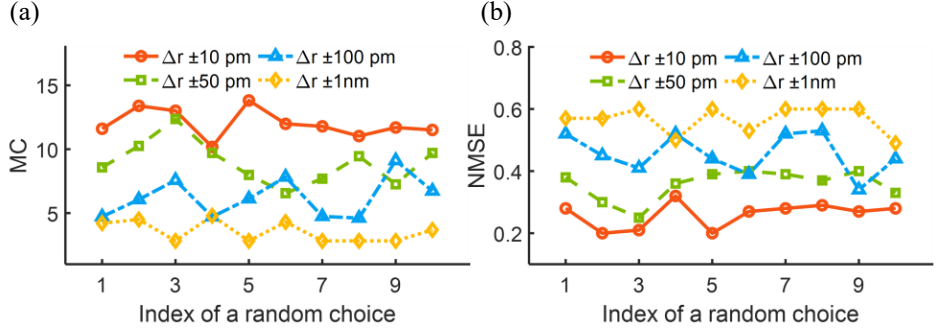

**Fig. 7.** Performance of Ten Sets of Linear Cavities within the single SCMRRs-10 system on the NARMA 10 Task. (a) MC and (b) NMSE versus the index of a random choice in different error range. Operated at $P_M = 0.1mW$ $\Delta\lambda_s = -10pm, Q_2/Q_1 = 300$.

## 6. Conclusion

In this paper, we conducted numerical investigation of a series-coupled MRRs system as a versatile computational platform in time-delay RC. Compared with the previous work based on waveguides of tens of centimeters [38], our scheme introduces series-coupled linear cavity array with micrometer scale footprint to provide enough MC. For on-chip waveguides with length of tens centimeters, not only the fabrication is challenging, but also the optical loss must be considered. Our scheme thus holds higher practicality and scalability. To evaluate its computational performance, we computed three typical tasks that have different memory requirements. For the NARMA 10 task, the proposed system achieves better performance than the MRR with optical feedback due to the great MC provided by these series-coupled MRRs. For the Mackey-Glass prediction task, because of the satisfaction of the system nonlinearity and the linear MC, the proposed system obtains almost the same lowest prediction error as the MRR with optical feedback. Ultimately, because the Santa Fe task does not need too much MC, the proposed system achieves a little better performance than the MRR with optical feedback [38]. All simulation results are calculated under the same bit period ($\tau = 1ns$) and the same number of virtual nodes (N=25). The proposed SCMRRs-based system exhibited nearly the same computational properties as the MRR with optical feedback, but with significantly smaller footprint. With existing fabrication techniques, this proposed system provides a route to scalable photonic RC based integrated chips.

**Funding**

National Natural Science Foundation of China (NSFC) (60907032, 61675183, 61675184); Natural Science Foundation of Zhejiang Province (LY16F050009, LY20F050009); Open Fund of the State Key Laboratory of Advanced Optical Communication Systems and Networks, China (2020GZKF013). Horizontal projects of public institution ( KY-H-20221007.KYY-HX-20210893).

**Acknowledgment.** This work was partially carried out at the USTC Center for Micro and Nanoscale Research and Fabrication. The authors thank Dr. G. Donati (IFISC institute for cross disciplinary physics and complex systems (CSIC-UIB), Spain) for the fruitful discussions.

**Disclosures.** The authors declare no conflicts of interest.